\documentclass[11pt,draftcls,onecolumn]{IEEEtran}
\usepackage{amssymb,amsmath,graphicx}

\newcommand{\FF}{\mathbb{F}}
\newcommand{\ZZ}{\mathbb{Z}}
\newcommand{\QQ}{\mathbb{Q}}

\newcommand{\Ac}{\mathcal{A}}
\newcommand{\Cc}{\mathcal{C}}
\newcommand{\Gc}{\mathcal{G}}
\newcommand{\Ic}{\mathcal{I}}
\newcommand{\Oc}{\mathcal{O}}
\newcommand{\Mc}{\mathcal{M}}
\newcommand{\Sc}{\mathcal{S}}

\newcommand{\xm}{\mathbf{X}}

\newcommand{\zom}{\mathbf{0}}
\newcommand{\id}{\mathbf{I}}

\newcommand{\af}{\mathfrak{a}}

\newcommand{\End}{{\rm End}}
\newcommand{\Gal}{{\rm{Gal}}}

\newcommand{\dmin}{d_{\rm min}}
\newcommand{\rnorm}{\rho_{\rm norm}}

\newtheorem{ex}{Example}
\newtheorem{lemma}{Lemma}
\newtheorem{defn}{Definition}
\newtheorem{prop}{Proposition}
\newtheorem{cor}{Corollary}

\title{
Codes over Matrix Rings \\ for Space-Time Coded Modulations}
\author{Fr\'ed\'erique Oggier, Patrick Sol\'e, Jean-Claude Belfiore
\thanks{F. Oggier is with Division of Mathematical Sciences, School of
Physical and Mathematical Sciences, Nanyang Technological University,
Singapore. P. Sol\'e and J.-C. Belfiore are with Telecom ParisTech, CNRS, UMR 5141, France.
Email:frederique@ntu.edu.sg,\{sole,belfiore\}@telecom-paristech.fr.
Part of this work appeared at ISIT 09~\cite{OSB09} and at ITW 09~\cite{OS09}.
}}

\begin{document}
\maketitle

\begin{abstract}
It is known that, for transmission over quasi-static MIMO fading channels with $n$ transmit antennas, diversity can be obtained by using an inner
fully diverse space-time block code while coding gain, derived from the determinant criterion, comes from an appropriate outer code.
When the inner code has a cyclic algebra structure over a number field, as for perfect
space-time codes, an outer code can be designed via coset coding. More precisely, we take the quotient of the algebra
by a two-sided ideal which leads to a finite alphabet for the outer code, with a
cyclic algebra structure over a finite field or a finite ring. We show that the determinant criterion induces various metrics
on the outer code, such as the Hamming and Bachoc distances. When $n=2$, partitioning the $2\times 2$
Golden code by using an ideal above the prime 2 leads to consider codes over either $\Mc_2(\FF_2)$ or $\Mc_2(\FF_2[i])$, both
being non-commutative alphabets.
Matrix rings of higher dimension, suitable for $3\times 3$ and $4\times 4$
perfect codes, give rise to more complex examples.

\textit{Keywords:} Space-time codes, codes over rings, cyclic algebras, number fields, finite
rings, Golden code.
\end{abstract}

%**************************************************************************%
%
% INTRO
%
%**************************************************************************%

\section{Introduction}
\label{sec:intro}

We consider the problem of coding over a quasi-static (slow) fading MIMO
channel, for example in a mobile wireless setting, where the channel is
assumed to be fixed over the duration of a frame. Compared to standard
MIMO channels, slow fading induces a loss in diversity, which can be
compensated by using concatenated coding schemes, as for example
space-time trellis codes \cite{TSC98}. Finer concatenated schemes enable
to distinguish the two main design criteria, namely the rank and determinant
criteria: an inner code guarantees full diversity, while combining with
an outer code brings coding gain. Any fully diverse space-time code can
be used as inner code, but in this work, we will focus on codes built over
cyclic division algebras \cite{SRS03,ORBV06} whose algebraic structure
is easier to analyze.

%***********************************************************************%
\subsection{Related work}

Most attempts in the literature to obtain coded modulation schemes for
algebraic space-time codes focused on having the so-called Golden code
\cite{BRV05} as inner code. In the first attempt \cite{CBRV05}, the Golden code
was concatenated with an outer trellis code, whose drawback is its high
trellis complexity. Trellis coded modulation using a set partitioning of the
Golden code is studied in \cite{HVB07}, where a systematic design approach is
proposed: partitions of the Golden code with increasing minimum determinant
correspond to $\ZZ^8$ lattice partitions, which are labeled by using a
sequence of nested binary codes. In \cite{LRBV08}, the algebraic structure of
the Golden code partitions is investigated, and the authors show that they
are actually dealing with matrices over the finite field $\FF_2$, or over
the finite ring $\FF_2[i]$. The problem becomes thus the one of designing a
suitable outer code over the given ring of matrices, for which only two
examples are given: one repetition code of length 2, and one ad hoc
construction using Reed Solomon codes. In \cite{OSB09}, codes over
$\Mc_2(\FF_2)$ and $\Mc_2(\FF_2[i])$ have been proposed, with applications
to modulation schemes for the Golden code.

Generalizations to higher dimensional perfect codes are reported in
\cite{SRB08} where a partition of a $4\times 4$ perfect code is considered,
and in \cite{OS09}, where it has been shown that for dimensions 3 and 4,
codes to be designed are over respectively $\Mc_3(\FF_4)$ and $\Mc_4(\FF_2)$.

%**************************************************************************%

\subsection{Contribution and organization}
The original motivation for this paper is the observation that
all previous works base their code design on a coarse bound
depending on the minimum Hamming distance of the outer code.
Instead, the determinant criterion drives us, here, to consider other weights than the
Hamming weight. These alternative weights can take more than
one nonzero value, a feature that allows us to derive finer
lower bounds.
Furthermore, the present paper deepens previous results in two main
ways. First, we explore outer code constructions when the inner code has higher dimension
than the Golden code: we propose, for instance, a multilevel code construction
over $\Mc_4(\FF_2)$.  Second, for $n=2$, we go one level deeper in the partition of the Golden code, by
quotienting with an ideal of higher norm. This
enlarges the base ring, thus moving matrix entries from $\FF_2$ to $\FF_2[i]$.

The material is organized in the following way. Section \ref{sec:coset} gives a general framework
for dealing with $n$-dimensional coset codes. It gives a sequence of isomorphisms yielding four
different representations of the outer code alphabet: the quotient of the inner code by
a two-sided ideal, an algebra of matrices over a finite field, a cyclic algebra over a finite field,
Cartesian products of finite fields. Section \ref{sec:codehigh} studies
weights on the outer code in relation with determinantal lower bounds, for $n=2,3,4$ and presents a
multilevel construction for $n=4$.
Section \ref{sec:M2F2} is dedicated to the special case $\Mc_2(\FF_2)$, where codes for
both the Hamming and Bachoc distances are considered.
In Section \ref{sec:M2F2i}, we extend the Bachoc weight to $\Mc_2(\FF_2[i])$ where a bidimensional
Lee-like distance is derived. Corresponding codes are proposed. Section
\ref{sec:summary} puts the preceding results into perspective and points out some
challenging open problems.

%**************************************************************************%
%
% Mn(F2^k)
%
%**************************************************************************%

\section{Coset codes}\label{sec:coset}

%***************************************************************************%

\subsection{Background}

For a slow block fading channel, where the fading coefficients
are assumed to be constant for $L$ time blocks, the goal is to design
a codebook $\tilde{\Cc}$ of codewords
\[
\xm=(X_1,\ldots,X_L),~X_i\in\Sc
\]
for $i=1,\ldots,L$, where $\Sc$ is a set of codewords from a fully
diverse space-time codebook, such that the minimum determinant $\Delta_{min}$
of $\tilde{\Cc}$, given by
\begin{eqnarray}
\Delta_{min}&=&\min_{\xm\neq{\bf 0}}\det(\xm\xm^*)\label{eq:detmin}\\
           &=&\min_{\xm\neq\zom} \det(X_1X_1^*+\ldots+X_LX_L^*)\nonumber\\
           &\geq&\min_{\xm\neq\zom} \left( \sum_{i=1}^L |\det(X_i)|\right)^2
            \label{eq:bounddetmin}
\end{eqnarray}
is maximized. In this paper, $\Sc$ will be a set of
$n\times n$ perfect space-time codewords \cite{ORBV06,ESV07}. These codes
are not only fully diverse, they further offer a good minimum determinant,
independently of the size of the signal constellation.

It is known that choosing the blocks $X_i$ independently
does not bring coding gain. This is remedied by using outer codes, or
more particularly in this setting, coset codes, as proposed in \cite{LRBV08}.
Consider the projection
\begin{eqnarray}
\pi: & \Sc & \rightarrow  \Sc/\Ic \simeq R \label{eq:proj}\\
     & X   & \mapsto \pi(X)\nonumber
\end{eqnarray}
where $\Ic$ is a two-sided ideal of $\Sc$ seen as a ring, so that the
quotient $\Sc/\Ic \simeq R$ is a ring.
We now take a code $\Cc$ over $R$. The coset code
$\tilde{\Cc}$ is obtained by considering $\pi^{-1}(\Cc)$.

To evaluate the spectral efficiency of $\tilde{\Cc}$ independently of the 
size of the signal constellation in use, we employ the notion of normalized 
redundancy $\rnorm$ per channel use, defined by
\begin{equation}\label{eq:rnorm}
\rnorm=\frac{\mbox{outer code redundancy bits}}{Ln}.
\end{equation}   

To build coset codes as described above,
the first step is to identify the quotient ring $\Sc/\Ic \simeq R$.
In this section, we show that if we start with $\Sc$
a code built over a cyclic algebra, then $\Sc/\Ic$ also has a cyclic algebra
structure however over a finite field.

%******************************************************************************%

\subsection{Cyclic algebras}

Let us briefly recall the definition of codes built over cyclic algebras,
introduced in \cite{SRS03}, since perfect space-time codes that are of
interest for this work are a subclass.
\begin{defn}\label{def:ca}
Let $L/K$ be a cyclic extension of degree $n$, with Galois group
$\Gal(L/K)=\langle\sigma\rangle$, where $\sigma$ is the generator of the cyclic
group. Let $\Ac=(L/K,\sigma,\gamma)$ be its corresponding
{\em cyclic algebra} of degree $n$, that is
\[
\Ac = 1L \oplus eL \oplus \ldots \oplus e^{n-1} L
\]
with $e\in\Ac$ such that $l e = e \sigma(l)$ for all $l\in L$
and $e^{n}=\gamma\in K$, $\gamma\neq 0$.
\end{defn}
Note that $L/K$ is a priori any cyclic field extension. In this paper, we
use both cyclic algebras over number fields and cyclic algebras over finite fields.

One can associate a matrix to any element $x\in\Ac$ using
the map $\lambda_{x}$, the multiplication by $x$ of an element $y\in\Ac$:
\[
\begin{array}{rcl}
\lambda_{x}:\Ac & \rightarrow & \Ac\\
y & \mapsto & \lambda_{x}(y)=x\cdot y.
\end{array}
\]
The matrix of the multiplication by $\lambda_{x}$, with
\[
x=x_0+ e x_1 +\ldots+ e^{n-1} x_{n-1},
\]
is given by
\begin{equation}
\left(\begin{array}{ccccc}
x_0 & \gamma\sigma(x_{n-1}) & \gamma\sigma^2(x_{n-2}) & \ldots & \gamma\sigma^{n-1}(x_1)    \\
x_1 & \sigma(x_0)           & \gamma\sigma^2(x_{n-1}) & \ldots & \gamma\sigma^{n-1}(x_2) \\
\vdots  &                   & \vdots                  &        & \vdots                  \\
x_{n-2} & \sigma(x_{n-3})   & \sigma^2(x_{n-4})       & \ldots & \gamma\sigma^{n-2}(x_{n-1})\\
x_{n-1} & \sigma(x_{n-2}) & \sigma^2(x_{n-3})  & \ldots & \sigma^{n-1}(x_0)
\end{array}\right).
\label{eq:matrix}
\end{equation}
Perfect codes \cite{ORBV06} are codes built over cyclic division algebras,
with in particular the property that their minimum determinant is lower
bounded by a constant independent of the size of the signal constellation.
This can be achieved by considering the subset of elements
$x=x_0+ e x_1 +\ldots+ e^{n-1} x_{n-1}$, $x_i$ in $\Oc_L$ instead of $L$,
$k=1,\ldots,n$, where $\Oc_L$ denote the ring of integers of $L$. In other
words, we consider the subset $\Lambda\subset\Ac$ given by
\[
\Lambda=1\Oc_L\oplus e\Oc_L\oplus\ldots \oplus e^{n-1}\Oc_L,
\]
which is actually an order of  $\Ac$, as identified in \cite{HLRV}.

For the case of interest to us, $K$ is typically $\QQ(i)$ or $\QQ(\zeta_3)$,
where $\zeta_3$ is a primitive third root of unity, to allow the use of
either QAM or HEX symbols. Since their respective rings of integers $\ZZ[i]$
and $\ZZ[\zeta_3]$ are principal ideal domains, it makes sense to speak of
an $\Oc_K$-basis for $\Oc_L$. We can now be more precise, and recall that
for $R$ a Noetherian integral domain with quotient field $K$, and $\Ac$ a
finite dimensional $K$-algebra, we have the following definition.
\begin{defn}\label{def:order}
An $R$-{\em order} in the $K$-algebra $\Ac$ is a subring $\Lambda$ of $\Ac$,
having the same identity element as $\Ac$, and such that $\Lambda$ is
a finitely generated module over $R$ and generates $\Ac$ as a linear space
over $K$. An order $\Lambda$ is called {\em maximal} if it is not properly
contained in any other $R$-order.
\end{defn}

In the cyclic algebra $\Ac$, we can choose the elements
$0\neq \gamma\in K$ to be an algebraic integer. We see that the order 
$\Lambda$ given above is more precisely an $\Oc_K$-order in $\Ac$.
Orders $\Lambda$ corresponding to the codes from \cite{ORBV06}, for dimensions 2,3 and 4, are reported in Table~\ref{table:order}.
\begin{table}
\begin{center}
\begin{tabular}{ccc}
\hline
$n$ & $L/K$ & $\Oc_L$ \\
\hline
2&$\QQ(i,\sqrt{5})/\QQ(i)$ & $\ZZ[i,(1+\sqrt{5})/2]$ \\
3&$\QQ(\zeta_3,\zeta_7+\zeta_7^{-1})/\QQ(\zeta_3)$&$\ZZ[\zeta_3,\zeta_7+\zeta_7^{-1}]$ \\
4&$\QQ(i,\zeta_{15}+\zeta_{15}^{-1})/\QQ(i)$&$\ZZ[i,\zeta_{15}+\zeta_{15}^{-1}]$\\
\hline
\end{tabular}
\caption{Orders corresponding to some perfect codes}
\label{table:order}
\end{center}
\end{table}
The table reads that for an $n\times n$ space-time block code, the cyclic
field extension used to construct the cyclic algebra $\Ac$ is $L/K$, and
the order $\Lambda$ in $\Ac$ is given by
$\Lambda=1\Oc_L\oplus e\Oc_L\oplus\ldots \oplus e^{n-1}\Oc_L$.
When $\Sc$ is a set of codewords coming from division
algebras, we can really consider $\Lambda$, an order of the algebra as in Definition \ref{def:order}, which has a ring structure.

The $\Oc_K$-order $\Lambda$ of $\Ac$ is a free module over $\Oc_K=\ZZ[i]$ or
$\ZZ[\zeta_3]$, with basis $\{b_i\}$, $i=1,\ldots,n^2$:
\[
\Lambda \simeq \bigoplus_{i=1}^{n^2}b_i\Oc_K,
\]
since for us $\Oc_L$ is a free $\Oc_K$-module of rank $n$ (say
with basis $\beta_k$, $k=1,\ldots,n$):
\begin{eqnarray*}
\Lambda &\simeq& \bigoplus_{j=1}^n e^j\Oc_L \\
        &\simeq& \bigoplus_{j=1}^n e^j \bigoplus_{k=1}^n \beta_k\Oc_K.
\end{eqnarray*}
The basis vectors $\{b_i\}$ are thus given by
\[
\{e^j\beta_k\},~j,k=1,\ldots,n.
\]

Let $\af$ be a two-sided ideal of $\Oc_K$. Since $\Oc_K$ is commutative,
we have that
\[
\Lambda/\af\Lambda \simeq \bigoplus_{i=1}^{n^2}b_i\Oc_K/b_i\af
\]
is a free module over the ring $\Oc_K/\af\Oc_K$, with basis
$\{\pi(b_i)\}$, $i=1,\ldots,n^2$, where $\pi$ is the canonical projection
\[
\pi:\Lambda\rightarrow \Lambda/\af\Lambda.
\]
The above considerations mean the following for our setting.
\begin{lemma}
For $n=2,3,4$ respectively, we have:
\begin{enumerate}
\item
If $\Lambda=\ZZ[i,(1+\sqrt{5})/2]+e\ZZ[i,(1+\sqrt{5})/2]$, $\af=(1+i)$, then
\[
\ZZ[i]/\af\ZZ[i] \simeq \FF_2
\]
and $\Lambda/\af\Lambda$ is a $\FF_2$-module of rank 4. In particular, we
have that
\[
|\Lambda/\af\Lambda|=2^4.
\]
\item
If $\Lambda=\ZZ[\zeta_3,\zeta_7+\zeta_7^{-1}]+e\ZZ[\zeta_3,\zeta_7+\zeta_7^{-1}]+
e^2\ZZ[\zeta_3,\zeta_7+\zeta_7^{-1}]$, $\af=2$, then
\[
\ZZ[\zeta_3]/\af\ZZ[\zeta_3] \simeq \FF_2^2
\]
and $\Lambda/\af\Lambda$ is a $\FF_4$-module of rank 9. In particular, we
have that
\[
|\Lambda/\af\Lambda|=4^9.
\]
\item
If $\Lambda=\ZZ[i,\zeta_{15}+\zeta_{15}^{-1}]+e\ZZ[i,\zeta_{15}+\zeta_{15}^{-1}]
+e^2\ZZ[i,\zeta_{15}+\zeta_{15}^{-1}]+e^3\ZZ[i,\zeta_{15}+\zeta_{15}^{-1}]$,
$\af=(1+i)$, then
\[
\ZZ[i]/\af\ZZ[i] \simeq \FF_2
\]
and $\Lambda/\af\Lambda$ is a $\FF_2$-module of rank 16. In particular,
we have that
\[
|\Lambda/\af\Lambda|=2^{16}.
\]
\end{enumerate}
\end{lemma}

Note that $2$ is prime in $\ZZ[\zeta_3]$, while $2$ ramifies as
$2=(1+i)(1-i)=(1+i)^2$ in $\ZZ[i]$.

This yields that
\begin{prop}
We have
\begin{eqnarray*}
\Lambda/\af\Lambda &\simeq& \Mc_n(\Oc_K/\af\Oc_K) \\
 &\simeq&
 \left\{
 \begin{array}{ll}
 \Mc_2(\FF_2) & \mbox{ for }n=2 \\
 \Mc_3(\FF_4) & \mbox{ for }n=3 \\
 \Mc_4(\FF_2) & \mbox{ for }n=4
 \end{array}
 \right.
\end{eqnarray*}
\end{prop}
\begin{IEEEproof}
By the previous lemma, we already know that $\Lambda/\af\Lambda$ is a
$\Oc_K/\af\Oc_K$-module whose cardinality is the same as
$\Mc_n(\Oc_K/\af\Oc_K)$. It is thus enough to give a ring homomorphism
$\psi:\Lambda/\af\Lambda\rightarrow \Mc_n(\Oc_K/\af\Oc_K)$ which is
one-to-one to conclude, and $\psi$ can be defined by mapping the basis vectors
$\pi(e^j\beta^k)$.
\end{IEEEproof}

The particular case for $n=2$ was proved in \cite{LRBV08}. The meaning
of this proposition is that when considering the projection (\ref{eq:proj})
\begin{eqnarray*}
\pi: & \Sc & \rightarrow  \Sc/\Ic \simeq R \\
     & X   & \mapsto \pi(X)
\end{eqnarray*}
to build coset codes with $\Sc$ coming from perfect codes, we need to build
codes over matrices over finite fields.
We prove next that an alternative point of view
is to ask for codes over cyclic algebras over finite fields.

%********************************************************************%

\subsection{Cyclic algebras over finite fields}\label{subsec:ffcycalg}

Let $\FF_2$ be the finite field with 2 elements, and consider
the field extension $\FF_{2^n}/\FF_2$ of degree $n$, that is
$\FF_{2^n}\simeq \FF_2(w)$ with $p(w)=0$ and $p \in \FF_2[X]$ is
an irreducible polynomial of degree $n$. Its cyclic Galois group
is generated by the Frobenius automorphism $\sigma:w\mapsto w^2$.
We consider the cyclic algebra $\Ac=(\FF_{2^n}/\FF_2,\sigma,1)$,
with
\[
\Ac \simeq \FF_{2^n} \oplus \ldots e\FF_{2^n}\oplus e^{n-1}\FF_{2^n}
\]
(see Definition \ref{def:ca}).
We know by Lemma 2.16 in \cite{S85} that $\Ac\simeq \End_{\FF_2}(\FF_{2^n})$.
The isomorphism $j:\Ac \rightarrow \End_{\FF_2}(\FF_{2^n})$ is explicitly
given by $j(a)$, which is the multiplication by $a$ for all $a$ in $\FF_{2^n}$,
and $j(e)=\sigma$. Indeed, we have that
\[
j(ae)(x) = (j(a)j(e))(x) = j(a)\sigma(x) = a\sigma(x)
\]
which in turn can be written
\[
j(e)(\sigma(a)x)= j(e)j(\sigma(a))(x)= j(e\sigma(a))(x)
\]
thus $j(ae)=j(e\sigma(a))$.

\begin{ex}\rm
We consider the cyclic algebra $\Ac=(\FF_8/\FF_2,\sigma,1)$, where
$\FF_8\simeq \FF_2(w)$ with $w^3+w+1=0$ and $\sigma:w\mapsto w^2$.
As a vector space, we have $\Ac \simeq \FF_8 \oplus e\FF_8 \oplus e^2\FF_8 $
and multiplication is given by $ae=e\sigma(a)$ for $a\in\FF_8$.
We have that $\Ac \simeq \Mc_3(\FF_2)$.
The isomorphism is given as follows:
\begin{eqnarray*}
e & \mapsto &
\left(
\begin{array}{ccc}
1 & 0 & 0\\
0 & 0 & 1 \\
0 & 1 & 1
\end{array}
\right)\\
a_0+a_1w+a_2w^2 &\mapsto&
\left(
\begin{array}{ccc}
a_0 & a_1 & a_2 \\
a_2 & a_0+a_2 &a_1\\
a_1 & a_1+a_2& a_0+a_2
\end{array}
\right).
\end{eqnarray*}
It is a straightforward computation to check that
\begin{eqnarray*}
(a_0+a_1w+a_2w^2)e&=& e\sigma(a_0+a_1w+a_2w^2)\\
                  &=& e(a_0+a_1w^2+a_2(w^2+w)).
\end{eqnarray*}
\end{ex}

\begin{ex}\rm
Consider now the cyclic algebra $\Ac=(\FF_{16}/\FF_2,\sigma,1)$, where
$\FF_{16}\simeq \FF_2(w)$ with $w^4+w^2+1=0$ and $\sigma:w\mapsto w^2$.
We have that $\Ac \simeq \Mc_4(\FF_2)$.
The isomorphism is given as follows:
\[
e  \mapsto
\left(
\begin{array}{ccccc}
1 & 0 & 0 & 0\\
0 & 0 & 1 & 0\\
1 & 1 & 0 & 0\\
0 & 0 & 1 & 1
\end{array}
\right),~
w \mapsto
\left(
\begin{array}{ccccc}
0 & 1 & 0 & 0 \\
0 & 0 & 1 & 0\\
0 & 0 & 0 & 1 \\
1 & 1 & 0 & 0
\end{array}
\right).
\]
\end{ex}

The above example gives us an explicit isomorphism
\begin{equation}\label{eq:M4}
\Mc_4(\FF_2) \simeq \FF_{16}\oplus e\FF_{16}\oplus e^2\FF_{16}\oplus e^3\FF_{16}.
\end{equation}

We now finish this sequence of isomorphisms, and connect codes on
cyclic algebras over finite fields to classical error correcting codes.
The isomorphism
\[
\Ac \simeq \FF_{2^n} \oplus \ldots \FF_{2^n}e\oplus\FF_{2^n} e^{n-1} \simeq \Mc_n(\FF_2)
\]
clearly induces an isomorphism of $\FF_2$-left vector space
\[
\phi: \underbrace{\FF_{2^n}\times \ldots \times \FF_{2^n}}_{n}\rightarrow \Mc_n(\FF_2).
\]
Also, $\phi$ can be extended to $L$-tuples
\begin{eqnarray*}
\phi:& (\FF_{2^n}\times\ldots\times \FF_{2^n})^L &\rightarrow  \Mc_n(\FF_2)^L
\end{eqnarray*}
so that if $\Cc$ is a code of length $L$ over $\Mc_n(\FF_2)$, then
$\phi^{-1}(\Cc)$ is a code of length $2L$ over $\FF_{2^n}$.

This connection with classical codes has been introduced first in \cite{B96}
for the construction of particular lattices.

%***********************************************************************%
%
% Error correcting codes
%
%***********************************************************************%

\section{Weights and Codes}\label{sec:codehigh}

In this section, we propose a multilevel construction for codes over
rings of matrices with coefficients in finite fields.
To see which performance the code should reach, we first compute a
bound on the minimum determinant.

%*************************************************************************%
\subsection{Hamming distance bound}

The determinant of a $n\times n$ codeword $X\in\Sc$ can be bounded depending on
its projection $\pi(X)\in \Sc/\Ic$, as follows for the case when
$\Ic=a\Sc$, $a$ a scalar.
\begin{lemma}\label{lem:det}
We have that
\begin{enumerate}
\item
$|\det(X)|^2\geq |a^n|^2\delta$ if $\pi(X)=\zom$, $\zom\neq X$,
\item
$|\det(X)|^2 \geq |a|^2\delta$ if $\zom\neq\pi(X)$ is not a unit,
\item
$|\det(X)|^2 \geq \delta$  if $\pi(X)$ is a unit,
\end{enumerate}
where $\delta=\min_{X\in\Sc} |\det(X)|^2$.
\end{lemma}
\begin{IEEEproof}
\begin{enumerate}
\item
If $\pi(X)=\zom$, with $\zom\neq X$, then $X=a\tilde{X}\subset a\Sc$,
and $\det(X)=a^n\det(\tilde{X})$.
\item
If $\zom\neq\pi(X)$ is not a unit, then its determinant has to be zero, that
is $\det(\pi(X))=0$, implying that $\det(X)$ is a multiple of $a$.
\item
If $\pi(X)$ is a unit, then its determinant has to be one too.
\end{enumerate}
\end{IEEEproof}

We have from (\ref{eq:bounddetmin}) that
\[
\Delta_{min} \geq\min_{\xm\neq\zom} \left( \sum_{i=1}^L |\det(X_i)|\right)^2.
\]
For each of the $X_i$, we have that $\pi(X_i)$ may or not be $\zom$, and
if it is non-zero, it may or not be invertible.
Ideally, we would like to be able to distinguish these three cases, since
we have by Lemma \ref{lem:det} that
\begin{eqnarray*}
|\det(X)| & \geq & |a^n|\sqrt{\delta} \mbox{ if }\pi(X)=\zom,~X\neq\zom\\
|\det(X)| & \geq & |a|\sqrt{\delta} \mbox{ if }\zom\neq\pi(X)
\mbox{ is not a unit }\\
|\det(X)| & \geq & \sqrt{\delta} \mbox{ if }\pi(X)\mbox{ is a unit}.
\end{eqnarray*}
To start with, let us give a bound which only takes into account
zero and non-zero elements (the all zero elements case is treated afterwards).
Let $\dmin^H$ be the minimum Hamming distance of the code $\pi(\Cc)$, which
is as in the classical case the number of different components between any
two pairs of codewords. If not all $\pi(X_j)=\zom$, then by
definition of $\pi(\Cc)$, there are at least $\dmin^H$ terms such
that $\pi(X_j)\neq \zom$. We give those $X_j$ a weight of $\sqrt{\delta}$
(instead of either $|a|\sqrt{\delta}$ or $\sqrt{\delta}$). Thus
\begin{eqnarray*}
\Delta_{min}  &\geq& \min_{\xm\neq\zom} \left( \sum_{i=1}^L |\det(X_i)|\right)^2\\
              &\geq& (\dmin^H)^2\delta.
\end{eqnarray*}
Note that this is far from a tight bound, since we would like most of the weight to be given to codewords whose projection is either zero or a non-invertible element.

Since the case where all the $\pi(X_j)=0$ is not included above ($\dmin^H$
does not apply), we treat this case separately. Let us thus assume that
$\pi(X_j)=0$ for all $j=1,\ldots,L$. We then have $X_j=a\tilde{X}_j$, and
\begin{eqnarray*}
\Delta_{min}
&=& \min_{\xm\neq\zom}  \det(X_1X_1^*+\ldots+X_LX_L^*)\\
&=& \min_{\xm\neq\zom}  \det(|a|^2\id_n)\det(\tilde{X_1}\tilde{X_1^*}+\ldots
+\tilde{X_L}\tilde{X}_L^*)\\
&=& |a|^{2n}\min_{\xm\neq\zom}\det(\tilde{X_1}\tilde{X_1^*}
+\ldots+\tilde{X_L}\tilde{X}_L^*)\\
&\geq& |a|^{2n}\delta.
\end{eqnarray*}
Note that $\Delta_{min}$ is actually equal to $|a|^{2n}\delta$ if
$X=(a\tilde{X}_1,\zom,\ldots,\zom)$.
The problem is that the criterion $\pi(X)=\zom$ does not allow to
distinguish $X\neq\zom$ and $\zom\neq X\subset a\Sc$.

\begin{lemma}\label{lem:hambound}{\bf (Hamming distance bound)}
We have that
\[
\Delta_{min}\geq \min(|a|^{2n}\delta,(\dmin^H)^2\delta).
\]
\end{lemma}

It was proved in \cite{LRBV08} for $n=2$ (that is the Golden code) that
\[
\Delta_{min}\geq \min\left(4\delta,(\dmin^H)^2\delta\right),~\Ic=(1+i),
\]
where $\dmin^H$ is the minimum Hamming distance of the code over $\Mc_2(\FF_2)$,
and $\delta=1/5$ is the minimum determinant of the Golden code.
This above lemma tells us that
\[
\Delta_{min}\geq \min\left(64\delta',(\dmin^H)^2\delta'\right),~\Ic=(2),
\]
for $n=3$ over $\Mc_3(\FF_4)$, $\delta'=1/49$, and
\[
\Delta_{min}\geq \min\left(16\delta'',(\dmin^H)^2\delta''\right),~\Ic=(1+i),
\]
for $n=4$ over $\Mc_4(\FF_2)$ and $\delta''=1/1125$.

%*************************************************************************%
\subsection{Multilevel coding for $n=4$}

We know from (\ref{eq:M4}) that
\[
\Mc_4(\FF_2) \simeq \FF_{16}\oplus e\FF_{16}\oplus e^2\FF_{16}\oplus e^3\FF_{16}.
\]
Note that $(1+e)^4=(1+e^2)(1+e^2)=0$, showing that $1+e$ is nilpotent.
We set $f=1+e$, and do a change of basis to get
\[
\Mc_4(\FF_2) \simeq \FF_{16}\oplus f\FF_{16}\oplus f^2\FF_{16}\oplus f^3\FF_{16}.
\]
Via this isomorphism, we can write a matrix $X\in\Mc_4(\FF_2)$ as an
element $x\in \FF_{16}\oplus f\FF_{16}\oplus f^2\FF_{16}\oplus f^3\FF_{16}$,
given by
\[
x=x_0+fx_1+f^2x_2+f^3x_3.
\]
Note that if $x_0=0$, then $x$ is not invertible, since
$x=f(x_1+fx_2+f^2x_3)$, with $f$ nilpotent.

Consequently, to a codeword $(\pi(X_1),\ldots,\pi(X_L))\in \Mc_4(\FF_2)^L$ corresponds a vector
\[
(x_{10}+fx_{11}+f^2x_{12}+f^3x_{13},\ldots,
x_{L0}+fx_{L1}+f^2x_{L2}+f^3x_{L3}).
\]
The first level of coding is done using a $(L,k_1,d_1)$ code $C_1$ that maps
$k_1$ symbols to $x_{10},\ldots,x_{L0}$. Similarly, the $i$th level of coding uses a $(L,k_i,d_i)$ code $C_i$ that maps
$k_i$ symbols to $x_{0i},\ldots,x_{Li}$, $i=1,2,3$.

\begin{figure}
\begin{center}
\label{fig:m4f2}
\includegraphics[scale=0.7]{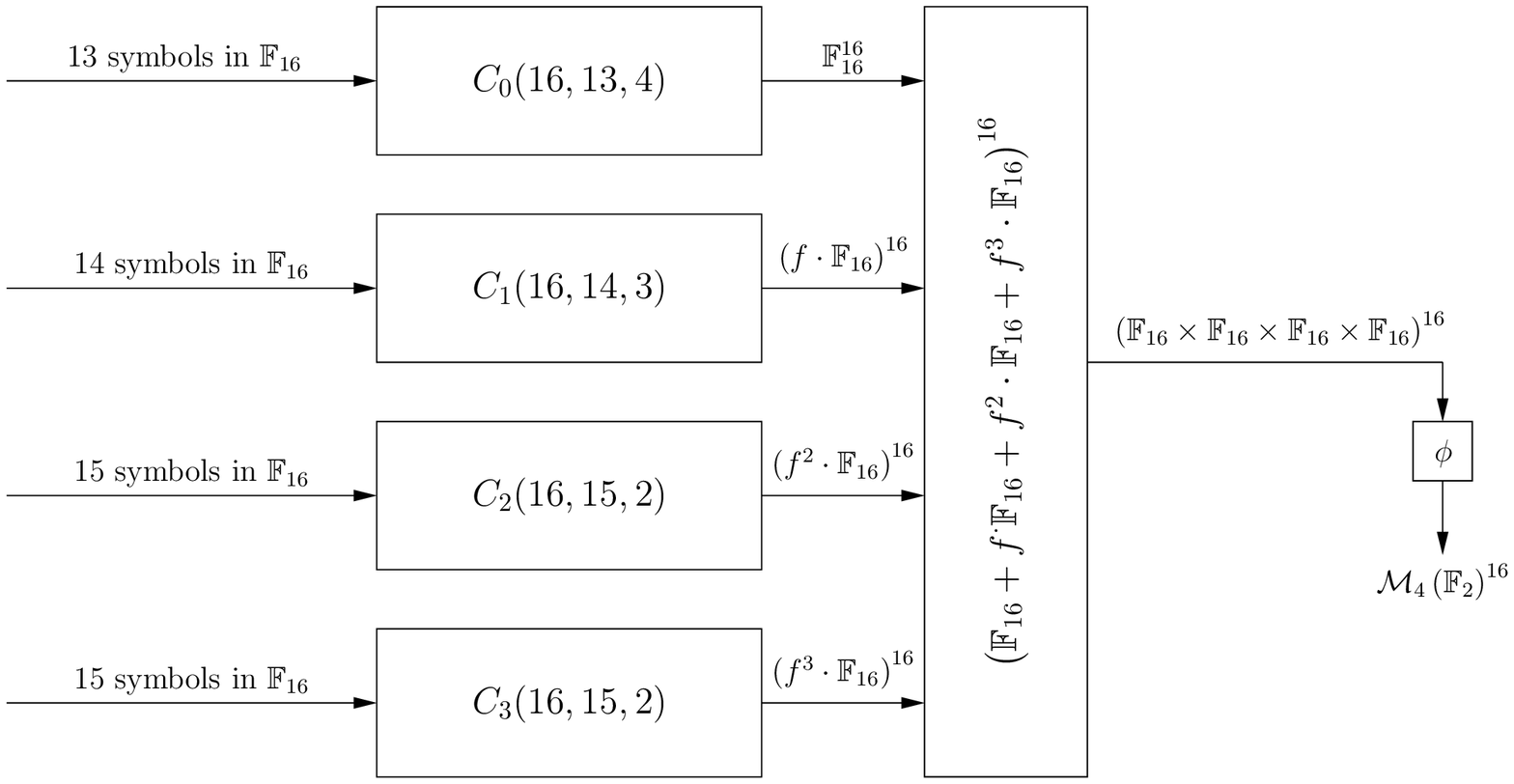}
\caption{Multilevel encoder for $\Mc_4(\FF_2)$.}
\end{center}
\end{figure}

Since $C_1$ has minimum distance $d_1$, either all coefficients are zero, or at least $d_1$ coefficients out of $x_{10},\ldots,x_{L0}$ are non-zero.
\begin{itemize}
\item
In the latter case, out of $(\pi(X_1),\ldots,\pi(X_L))$,  $d_1$ matrices may or may not be invertible, thus having a determinant that may or may not be invertible. An invertible determinant $\det(\pi(X))$ gives the lowest weight, that is,
$|\det(X)|^2\geq \delta$, and all we can say is
\[
\min \sum_{i=1}^L |\det(X_i)|\geq d_1 \sqrt{\delta}.
\]
\item
For $x_{10}=\ldots=x_{L0}=0$, we can do the same reasoning for $C_2$, except that the situation is more favorable: indeed, all matrices
$(\pi(X_1),\ldots,\pi(X_L))$ are now not invertible, meaning that their
determinant is a multiple of $1+i$, showing that in this case
\[
\min \sum_{i=1}^L |\det(X_i)|\geq \sqrt{2}d_2\sqrt{\delta}.
\]
\end{itemize}
By iterating the same steps for $C_2$ and $C_3$, we get that
\[
\min \sum_{i=1}^L |\det(X_i)|\geq
\min\{d_1,\sqrt{2}d_2,2d_3,2\sqrt{2}d_3\}\sqrt{\delta}.
\]

Note that this bound takes only into account the multilevel code, and
not the fact that we use a coset code, which gives a further constraint
(as shown in the previous subsection), finally yielding:
\[
\min \sum_{i=1}^L |\det(X_i)|\geq
\min\{4,d_1,\sqrt{2}d_2,2d_3,2\sqrt{2}d_3\}\sqrt{\delta}.
\]
It is thus enough to guarantee:
\[
d_1=4,~d_2=3,~d_3=2,~d_4=2.
\]
Parity codes can be used for $C_3$ and $C_4$.
For example, for $L=16$, we can choose\\
\begin{center}
\begin{tabular}{ccl}
$C_1$ &=& the $(16,13,4)$ Reed Solomon code,\\
$C_2$ &=& the $(16,14,3)$ Reed Solomon code,\\
$C_3$ &=& the $(16,15,2)$ parity check code,\\
$C_4$ &=& the $(16,15,2)$ parity check code,
\end{tabular}
\end{center}
for a rate of $47/64\approx0.734$. 
From (\ref{eq:rnorm}), the normalized redundancy of $\tilde{\Cc}$ when 
the above outer code $\Cc$ is used is
\begin{eqnarray*}
\rnorm&=&\frac{\mbox{outer code redundancy bits}}{Ln}\\
      &=&\frac{ (1+1+2+3\mbox{ symbols in }\FF_{16})(4\mbox{ bits})}{4L}\\
      &=& \frac{28}{64}=\frac{7}{16}.
\end{eqnarray*}

%**************************************************************************%
%
% M2(F2)
%
%**************************************************************************%

\section{Codes over $\Mc_2(\FF_2)$}
\label{sec:M2F2}

In the rest of the paper, we pay a special attention to the case $n=2$,
for which we take as inner code the Golden code
$\Gc$ \cite{BRV05}:
\begin{defn}\label{def:GC}
A codeword $X$ belonging to the Golden code $\Gc$ has the form
\[
X =
 \frac{1}{\sqrt{5}}\left(\begin{array}{cc}
\alpha(a+b\theta) & \alpha(c+d\theta)\\
i\bar{\alpha}(c+d\bar{\theta}) & \bar{\alpha}(a+b\bar{\theta})
\end{array}\right)
\]
where $a,b,c,d$ are QAM symbols (that is, $a,b,c,d\in\ZZ[i]$),
$\theta=\frac{1+\sqrt{5}}{2}$,
$\bar\theta=\frac{1-\sqrt{5}}{2}$, $\alpha=1+i-i\theta$ and
$\bar{\alpha}=1+i -i\bar{\theta}$.
Its minimum determinant is given by
\[
\delta=\min_{{\bf 0} \neq X \in\Gc}|\det(X)|^2
= \frac{1}{5},
\]
in particular it is always different from 0, and the Golden
code is fully diverse.
\end{defn}

The ring structure of the Golden code is best seen if we rewrite
\[
\Gc=\alpha(\ZZ[i,\theta]\oplus e\ZZ[i,\theta]),
\]
where $e$ is an element of $\Gc$ such that $e^2=i$, as already
mentioned in Table \ref{table:order}.
In what follows, we will see $\Gc$ either formally as above, or as
a set of matrices.

%*****************************************************************************%
\subsection{Codes over $\Mc_2(\FF_2)$ and Hamming weight}
\label{subsec:M2F2Ham}

We now discuss how codes over $\Mc_2(\FF_2)$ can be obtained from codes
over $\FF_4$.
We start by showing how error correcting codes over $\FF_4$ can be expressed
as codes over $\Mc_2(\FF_2)$. The starting point is the correspondence between
elements in $\FF_4$ and matrices in $\Mc_2(\FF_2)$ as given by the lemma below.

\begin{lemma}
Let $\FF_2$ be the finite field with 2 elements, and $\FF_4=\FF_2(\omega)$
be the finite field with 4 elements, where $\omega^2+\omega+1=0$.
There is a correspondence between the element $a=a_1+a_2\omega$ in $\FF_4$
and the matrix
\[
M_a=
\left(
\begin{array}{cc}
a_1 & a_2 \\
a_2 & a_1+a_2
\end{array}
\right).
\]
\end{lemma}
\begin{IEEEproof}
The matrix $M_a$ is just the multiplication matrix by $a$, since
\[
(1,\omega)
\left(
\begin{array}{cc}
a_1 & a_2 \\
a_2 & a_1+a_2
\end{array}
\right)
= (a,a\omega).
\]
\end{IEEEproof}

We thus define a code $\Cc$ over $\Mc_2(\FF_2)$ as follows:
let $(x_1,\ldots,x_L)$ be a codeword of an $[L,k,d]$ over $\FF_4$.
Then
\[
\Cc=\{ (X_1,\ldots,X_L)~|~X_i = M_{x_i} \in \Mc_2(\FF_2),~i=1,\ldots L\}.
\]
The code is clearly linear since $M_{a-b}=M_a-M_b$.

This allows to easily show that minimum distance $\dmin^H$ of the code over
$\Mc_2(\FF_2)$ is $d$, the minimum distance of the code over $\FF_4$.
Indeed, we have that
\[
\dmin=\min_{\zom\neq X}w_H((X_1,\ldots,X_L))
\]
and
\[
X_i=M_{x_i}=\zom \iff x_i=0.
\]

\begin{ex}\label{ex:dualrep}\rm
Consider the $[4,3,2]$ cyclic linear code over $\FF_4$, given by the dual
of the repetition code of length 4. Since the generator matrix of the
repetition code is $G=(1,1,1,1)$, its parity check matrix is thus
\[
H=\left(
\begin{array}{cccc}
1 & 1 & 0 & 0 \\
1 & 0 & 1 & 0 \\
1 & 0 & 0 & 1\\
\end{array}
\right)
\]
which is in turn the generator matrix of the dual.
Thus a codeword is of the form
\[
(x_1+x_2+x_3,x_1,x_2,x_3).
\]
In words, any coefficient is the sum of the 3 others, thus clearly any shift of
such codeword is also a codeword, and we obtain a parity code, which is cyclic, with parameters $[4,3,2]$.

The corresponding codeword over $\Mc_2(\FF_2)$ is
\[
X=(X_1+X_2+X_3,X_1,X_2,X_3)
\]
where
\[
X_i=
\left(
\begin{array}{cc}
x_{i1} & x_{i2} \\
x_{i2} & x_{i1}+x_{i2}
\end{array}
\right)\in\Mc_2(\FF_2)
\]
and
\[
x_i=x_{i1}+x_{i2}\omega,~i=1,2,3,~x_{i1},x_{i2}\in\FF_2.
\]
Thus
\[
\Cc=\{ (X_1+X_2+X_3,X_1,X_2,X_3)~|~X_i=M_{x_i}\in\Mc_2(\FF_2)\}
\]
is a linear code over $\Mc_2(\FF_2)$ with $\dmin^H=2$.
\end{ex}

Now we know from Lemma \ref{lem:hambound} that
\[
\Delta_{min}\geq \min(4\delta,(\dmin^H)^2\delta),
\]
so that the parity code is a good candidate, since it satisfies
\[
\dmin^H=2.
\]
From (\ref{eq:rnorm}), the normalized redundancy of $\tilde{\Cc}$ when 
the dual of the repetition code is used as outer code is
\begin{eqnarray*}
\rnorm&=&\frac{\mbox{outer code redundancy bits}}{Ln}\\
      &=& \frac{(L-1)+4}{2L}=\frac{L+3}{2L}.
\end{eqnarray*}

This code has the right minimum distance with respect to the bound of 
Lemma \ref{lem:hambound}, and is easily available for arbitrary values of $L$.  
However its normalized redundancy could be improved, which motivates a second construction.
Let
\[
\Cc_M =\{(X_1,X_2,\ldots,X_{L-1},X_1+X_2+\ldots+X_{L-1})~|~X_i\in\Mc_2(\FF_2)\}
\]
be a code defined directly over $\Mc_2(\FF_2)$ by mimicking our first
construction. It encodes $L-1$ elements of $\Mc_2(\FF_2)$ (for a total
of $4(L-1)$ bits) into a vector of length $L$. It is clearly linear, and
its minimum distance is $\dmin^H=2$. The normalized redundancy is now 
\begin{eqnarray*}
\rnorm&=&\frac{\mbox{outer code redundancy bits}}{Ln}\\
      &=& \frac{4}{2L}.
\end{eqnarray*}

\begin{table}
\begin{center}
\begin{tabular}{lccc}
\hline
code & coding gain & $\rnorm$ (bpc) & $L$ \\
\hline
parity code I  & $4/5$ & $\frac{L+3}{L}$ & arbitrary \\
parity code II & $4/5$ & $\frac{4}{L}$ & arbitrary\\
\hline
\end{tabular}
\label{table:summaryM2f2}
\caption{Summary of the performance of the proposed codes.}
\end{center}
\end{table}

%**************************************************************************%
\subsection{Codes over $\Mc_2(\FF_2)$ and Bachoc weight}
\label{subsec:M2F2Bac}

So far, we have provided code constructions based on the design criterion
of Lemma \ref{lem:hambound}, which is actually a coarse bound, as already
noticed during its derivation.  Recall from  Lemma \ref{lem:det} that
for $n=2$ and $\Ic=(1+i)$
\begin{eqnarray*}
|\det(X)| & \geq & 2\sqrt{\delta} \mbox{ if }\pi(X)=\zom,~X\neq\zom\\
|\det(X)| & \geq & \sqrt{2\delta} \mbox{ if }\zom\neq\pi(X)
\mbox{ is not a unit }\\
|\det(X)| & \geq & \sqrt{\delta} \mbox{ if }\pi(X)\mbox{ is a unit}.
\end{eqnarray*}
To get the bound of Lemma \ref{lem:hambound}, we use the Hamming weight,
that is, we assign a weight of either 1 or 0 on matrices in $\Mc_2(\FF_2)$,
to which corresponds a weight of $\sqrt{\delta}$ to each $X$ such that
$\pi(X)\neq \zom$, and a weight of zero otherwise. We are thus losing a
lot of information. In this section, we introduce a new weight to replace
the Hamming weight, which will tighten the bound for the minimum determinant.

We consider the new weight $w_B$ on $\Mc_2(\FF_2)$, that we call Bachoc
weight, as proposed by C. Bachoc in \cite{B96}, by setting
\begin{equation}\label{eq:wb}
w_B(Y)=
\left\{
\begin{array}{cl}
0 & Y=\zom\\
1 & Y \mbox{ is a unit } \\
2 & \zom\neq Y \mbox{ is not a unit}
\end{array}
\right..
\end{equation}
Correspondingly, we get for $X$ the weight
\[
\left\{
\begin{array}{cl}
0 & \mbox{ if } \pi(X)=\zom\\
\sqrt{\delta} & \mbox{ if } \pi(X) \mbox{ is a unit } \\
\sqrt{2\delta} &\mbox{ if } \zom\neq \pi(X) \mbox{ is not a unit}
\end{array}
\right..
\]

\begin{table}
\begin{center}
\begin{tabular}{cccc}
\hline
$\pi(X)$ & ideal weight     & Bachoc weight & Hamming weight \\
\hline
$\zom$   & $2\sqrt{\delta}$ & 0             & 0              \\
non-unit & $\sqrt{2\delta}$ & $\sqrt{2\delta}$& $\sqrt{\delta}$\\
unit     & $\sqrt{\delta}$  & $\sqrt{\delta}$ & $\sqrt{\delta}$\\
\hline
\end{tabular}
\caption{Comparison among weights for $\zom\neq X$, depending on $\pi(X)$}
\label{table:comp}
\end{center}
\end{table}

This weight is clearly finer than the Hamming weight (see Table
\ref{table:comp} for a comparison), since it allows to distinguish
non-zero invertible and non-invertible matrices in $\Mc_2(\FF_2)$.

\begin{defn}
We naturally define the corresponding minimum Bachoc distance
\[
\dmin^B=\min_{Y\neq Y'}w_B(Y-Y').
\]
\end{defn}

Let us now see how we can revisit the original design criterion based on the
new weight we have just introduced.
Recall that by (\ref{eq:bounddetmin})
\begin{eqnarray*}
\Delta_{min} &=& \min_{\xm\neq\zom}\det(\xm\xm^*) \\
             &\geq& \min_{\xm\neq\zom} \left( \sum_{i=1}^L |\det(X_i)|\right)^2.\\
\end{eqnarray*}
Let us now look at $\sum_{i=1}^L |\det(X_i)|$. It is lower bounded by
$(L+i(\sqrt{2}-1))\sqrt{\delta}$, if $i$ counts the number of non-invertible
projections. In particular, the lower bound ranges from $L$ to $\sqrt{2}L$.
Now the Bachoc weight, again if $i$ counts the number of non-invertible
projections, is $L+i$, which ranges from $L$ to $2L$. Thus
$\sum_{j=1}^L |\det(X_j)|\geq (L+i(\sqrt{2}-1))\sqrt{\delta}
\geq ((L+i)/\sqrt{2})\sqrt{\delta}$, and
\begin{eqnarray*}
\Delta_{min} &\geq & \min\left(\sqrt{\delta} w_B/\sqrt{2}\right)^2\\
             & = & \frac{(\delta\dmin^B)^2}{2}.
\end{eqnarray*}

\begin{lemma}\label{lem:bachbound}{\bf (Bachoc distance bound)}
We have that
\[
\Delta_{min}\geq \min\left(4\delta,\frac{(\dmin^B)^2}{2}\delta\right).
\]
\end{lemma}

Let us now see how to construct codes where the Bachoc weight can
be controlled.
Let again $\FF_2$ be the finite field with 2 elements, and
$\FF_4=\FF_2(\omega)$ be the finite field with 4 elements, where
$\omega^2+\omega+1=0$. As shown in Subsection \ref{subsec:ffcycalg}, we have a ring isomorphism
\begin{equation}\label{eq:iso1}
\Mc_2(\FF_2)\simeq\FF_2(\omega)+j\FF_2(\omega)
\end{equation}
where $j^2=1$ and $\omega j= j\omega^2$,
which is explicitly given by
\[
\left(
\begin{array}{cc}
0 & 1\\
1 & 0
\end{array}
\right) \mapsto j,~
\left(
\begin{array}{cc}
0 & 1\\
1 & 1
\end{array}
\right) \mapsto \omega.
\]
It in turn induces an isomorphism of $\FF_2$ left vector space
\[
\phi: \FF_4\times \FF_4\rightarrow \Mc_2(\FF_2).
\]
We have that $\phi$ maps a pair $(a,b)\in \FF_4\times \FF_4$ to a matrix
in $\Mc_2(\FF_2)$.
Since the elements $(a,0)$ and $(0,b)$ can be
identified with $a,bj \in \FF_4$ respectively, their image yields an invertible
matrix in $\Mc_2(\FF_2)$ whenever $a,b \in \FF_4^*$. These 6 elements thus
correspond to the 6 invertible matrices of $\Mc_2(\FF_2)$, which establishes
a one-to-one correspondence between elements of Hamming weight 1 in
$\FF_4^2$ and invertible matrices in $\Mc_2(\FF_2)$.

Furthermore, we have that $\phi$ is actually an {\em isometry}. This is a one
line proof, but due to its importance for our problem, let us repeat it as a
lemma.
\begin{lemma}\label{lem:isometry}
The map $\phi$ as defined above is an isometry.
\end{lemma}
\begin{IEEEproof}
We have
\[
w_B(Y)=w_B(\phi(y))=w(y)
\]
where $w$ denotes the Hamming weight.
\end{IEEEproof}

Let $d$ be the minimum Hamming distance of a code over $\FF_4$, that is
\[
d=\min_{0\neq x} w(x).
\]
It follows from the lemma that
\[
\dmin^B=\min_{{\bf 0}\neq X}w_B(X)=\min_{0\neq x}w_B(\phi(x))=d.
\]
Thus, to get a code over $\Mc_2(\FF_2)$ with a suitable Bachoc distance, it
is enough to construct a code over $\FF_4$ with the same minimum distance.
Since the code is brought back from $\FF_4$ via $\phi$, let us start by
being completely explicit:
\[
a+\omega b \mapsto
\left(
\begin{array}{cc}
a & b \\
b & a+b \\
\end{array}
\right),~a,b\in \FF_2,
\]
so that
\begin{eqnarray}
(a+b\omega,c+d\omega)
&\mapsto&
\left(
\begin{array}{cc}
a & b \\
b & a+b \\
\end{array}
\right)+
\left(
\begin{array}{cc}
c & d \\
d & c+d \\
\end{array}
\right)
\left(
\begin{array}{cc}
0 & 1 \\
1 & 0 \\
\end{array}
\right)\nonumber \\
&=&
\left(
\begin{array}{cc}
a+d & b+c \\
b+c+d & a+b+d \\
\end{array}
\right)\label{eq:iso}.
\end{eqnarray}

Vice-versa, we have that $\phi^{-1}$ is given by
\[
\left(
\begin{array}{cc}
y_{11} & y_{12} \\
y_{21} & y_{22} \\
\end{array}
\right)
\mapsto
( (y_{11}+y_{12}-y_{21})+\omega(-y_{11}+y_{22}),
(y_{11}+y_{12}-y_{22})+\omega(-y_{12}+y_{21}) ).
\]

Also, $\phi$ can be extended to $L$-tuples
\begin{eqnarray*}
\phi:& (\FF_4\times \FF_4)^L &\rightarrow  \Mc_2(\FF_2)^L
\end{eqnarray*}
so that if $\Cc$ is a code of length $L$ over $\Mc_2(\FF_2)$,
then $\phi^{-1}(\Cc)$ is a code of length $2L$ over $\FF_4$.

Let us now give two examples to illustrate both the map $\phi$ and $\phi^{-1}$.
\begin{ex}\label{ex:rep}\rm
Let us look at the repetition code of length 2.
We have that
\[
(Y,Y),
~Y=
\left(
\begin{array}{cc}
y_{11} & y_{12} \\
y_{21} & y_{22} \\
\end{array}
\right)
\in \Mc_2(\FF_2)
\]
is mapped to
\[
(y_1,y_2,y_1,y_2)\in \FF_4^4
\]
where
\begin{eqnarray*}
y_1&=& y_{11}+y_{12}-y_{21} +\omega(-y_{11}+y_{22}),\\
y_2&=& y_{11}+y_{12}-y_{22}+\omega(-y_{12}+y_{21}).
\end{eqnarray*}
We see that if $Y$ is invertible, then $w_B(Y)=1$, and $w_B((Y,Y))=2$.
On the other hand, $\phi^{-1}(Y)=(y,0)$, $0\neq y\in\FF_4$, so that
$\phi^{-1}((Y,Y))=(y,0,y,0)$, and $w((y,0,y,0))=2$.
Now, if $Y$ is not invertible, then $w_B(Y)=2$, and $w_B((Y,Y))=4$.
Furthermore $\phi^{-1}(Y)=(y_1,y_2)$, $0\neq y_1,y_2\in\FF_4$, so that
$\phi^{-1}((Y,Y))=(y_1,y_2,y_1,y_2)$, and $w((y_1,y_2,y_1,y_2))=4$.
Thus the minimum weight is given by
\[
\dmin^B=\min(2,4)=2=d,
\]
where $d$ is the minimum Hamming distance of the code over $\FF_4$.
Its normalized redundancy is from (\ref{eq:rnorm})
\[
\rnorm=\frac{4}{2L}=1.
\]
\end{ex}
\begin{ex}\label{ex:hex}\rm
Let us now consider the [6,3,4] hexacode, that is a linear code over $\FF_4$
of length 6, dimension 3, and minimum distance 4,
whose generator matrix is given by
\[
\left(
\begin{array}{cccccc}
1 & 0 & 0 & 1 & \omega & \omega \\
0 & 1 & 0 & \omega & 1 & \omega \\
0 & 0 & 1 & \omega & \omega & 1
\end{array}
\right).
\]
A codeword of the hexacode thus has the following form
\[
y=(y_1,y_2,y_3,y_1+\omega(y_2+y_3),y_2+\omega(y_1+y_3),y_3+\omega(y_1+y_2)).
\]
We now compute $\phi(y)$, using (\ref{eq:iso}). We have that
\begin{eqnarray*}
(y_1,y_2) &\mapsto& y_1+y_2j=(y_{11}+y_{12}\omega)+(y_{21}+y_{22}\omega)j \\
&\mapsto&
\left(
\begin{array}{cc}
y_{11} & y_{12} \\
y_{12} & y_{11}+y_{12} \\
\end{array}
\right)+
\left(
\begin{array}{cc}
y_{21} & y_{22} \\
y_{22} & y_{21}+y_{22} \\
\end{array}
\right)
\left(
\begin{array}{cc}
0 & 1 \\
1 & 0 \\
\end{array}
\right)\\
&=&Y_1,
\end{eqnarray*}
and
\begin{eqnarray*}
&&(y_3,y_1+\omega(y_2+y_3))\\
&\mapsto & y_3+[y_1+\omega(y_2+y_3)]j\\
&&=(y_{31}+y_{32}\omega)+[y_{11}+y_{22}+y_{32}+
\omega(y_{12}+y_{21}+y_{31}+y_{22}+y_{32})]j \\
&\mapsto &
\left(
\begin{array}{cc}
y_{31} & y_{32} \\
y_{32} & y_{31}+y_{32} \\
\end{array}
\right)+
\left(
\begin{array}{cc}
y_{11}+y_{22}+y_{32} & y_{12}+y_{21}+y_{31}+y_{22}+y_{32} \\
y_{12}+y_{21}+y_{31}+y_{22}+y_{32} & y_{11}+y_{12}+y_{21}+y_{31} \\
\end{array}
\right)
\left(
\begin{array}{cc}
0 & 1 \\
1 & 0 \\
\end{array}
\right)\\
&=& Y_2.
\end{eqnarray*}
A similar computation holds for $(y_2+\omega(y_1+y_3),y_3+\omega(y_1+y_2)) $
and yields $Y_3$.
Thus
\[
\phi(y)=(Y_1,Y_2,Y_3),
\]
a code of length 3 over $\Mc_2(\FF_2)$.
Since the hexacode has Hamming distance $d=4$, the minimum weight $\dmin^w$
of the code over $\Mc_2(\FF_2)$ is 4.

If we take for example $y_1=y_2=0$, we get that
\[
y=(0,0,y_3,\omega y_3,\omega y_3,y_3),
\]
and
\[
\phi(y)=\left( {\bf 0},~
\left(
\begin{array}{cc}
y_{32} & 0 \\
y_{31}+y_{32} & 0
\end{array}
\right),~
\left(
\begin{array}{cc}
0 & y_{32}  \\
0 & y_{31}+y_{32}
\end{array}
\right)
\right)
\]
which has weight 4, since we have two non-invertible matrices different from
${\bf 0}$.

Its normalized redundancy from (\ref{eq:rnorm}) is
\[
\rnorm=\frac{3\mbox{ symbols in }\FF_4}{2L}
=\frac{6}{6}=1.
\]
\end{ex}

The next example shows that the minimum Hamming distance and the minimum Bachoc
distance yield two different criteria: we will exhibit a code with minimum
Bachoc distance of 2, yet of minimum Hamming distance of 1.
\begin{ex}\rm
Consider again the $[4,3,2]$ code, the dual of the repetition
code as in Example \ref{ex:dualrep}, with generator matrix
\[
\left(
\begin{array}{cccc}
1 & 1 & 0 & 0 \\
1 & 0 & 1 & 0 \\
1 & 0 & 0 & 1\\
\end{array}
\right),
\]
so that a codeword is of the form
\[
(x_1+x_2+x_3,x_1,x_2,x_3),~x_i=x_{i1}+\omega x_{i2}\in\FF_4,
~x_{i1},x_{i2}\in\FF_2.
\]
Now
\[
\begin{array}{c}
\phi((x_1+x_2+x_3,x_1,x_2,x_3))=\\
\left(
\left(
\begin{array}{cc}
x_{11}+x_{21}+x_{31}+x_{12} & x_{12}+x_{22}+x_{32}+x_{11}\\
x_{22}+x_{32}+x_{11} & x_{11}+x_{21}+x_{31}+x_{22}+x_{32}
\end{array}
\right),
\left(
\begin{array}{cc}
\begin{array}{cc}
x_{21}+x_{32} & x_{22}+x_{31}\\
x_{22}+x_{32}+x_{31} & x_{21}+x_{22}+x_{32}
\end{array}
\end{array}
\right)
\right)
\end{array}.
\]
Consider the codeword
\[
(0,1,1,0),
\]
of Hamming weight 2, we have that
\[
\phi((0,1,1,0))=
\left(
\left(
\begin{array}{cc}
0 & 0\\
0 & 0
\end{array}
\right),
\left(
\begin{array}{cc}
\begin{array}{cc}
1 & 1\\
1 & 1
\end{array}
\end{array}
\right)
\right)
\]
which is of Hamming weight 1. It is of Bachoc weight 2 though,
since the non-zero matrix is not invertible.
\end{ex}

%**************************************************************************%
%
% M2(F2)
%
%**************************************************************************%

\section{Codes over $\Mc_2(\FF_2[i])$}
\label{sec:M2F2i}

The bottleneck for lower bounding the performance of coset codes is coming
from the codeword whose projection is all zero. In order to increase the
lower bound, one has to take a quotient by an ideal of higher norm.
This is the goal of this section.
Consider the projection
\[
\pi:\Gc \rightarrow \Gc/(2)\Gc \simeq \Mc_2(\FF_2[i]),~X \mapsto \pi(X),
\]
which maps a codeword $X$ in $\Gc$ to a matrix $\pi(X)$ in $\Mc_2(\FF_2[i])$.
Let $\FF_4=\FF_2(\omega)$ denote the finite field with 4 elements, where
$\omega^2+\omega+1=0$. Let us first note that the isomorphism
(\ref{eq:iso1}) can be easily extended:
\[
\Mc_2(\FF_2[i])\simeq\FF_2(\omega)[i]+j\FF_2(\omega)[i]\simeq\FF_4[i]+j\FF_4[i]
\]
where $j^2=1$ and $\omega j=j\omega^2$, and as before, it induces an
isomorphism
\[
\psi: \FF_4[i]\times\FF_4[i] \rightarrow \Mc_2(\FF_2[i]).
\]
We have that $\psi$ maps a pair $(a,b)\in  \FF_4[i]\times\FF_4[i]$ to a matrix
in $\Mc_2(\FF_2[i])$, as described in (\ref{eq:iso}), and $\psi$ can be
extended to $L$-tuples
\begin{eqnarray*}
\psi:& (\FF_4[i]\times\FF_4[i])^L &\rightarrow  \Mc_2(\FF_2[i])^L
\end{eqnarray*}
so that if $\Cc$ is a code of length $L$ over $\Mc_2(\FF_2[i])$,
then $\psi^{-1}(\Cc)$ is a code of length $2L$ over $\FF_4[i]$.

%************************************************************************%
\subsection{The structure of $\FF_4[i]$}

In the following, we may write an element $x\in\FF_4[i]$ as $x=a+b\omega$,
$a,b\in\FF_2[i]$, or $x=a'+b'i$, $a',b'\in\FF_4$, depending on the context.
The units of $\FF_4$ are as usual denoted by $\FF_4^*$.

The restriction $\phi$ of $\psi$ to $\FF_4\times\FF_4$ as been studied in
Subsection \ref{subsec:M2F2Bac}, where we noticed that $\phi$ maps a pair
$(a,b)\in \FF_4\times\FF_4 $ to a matrix in $\Mc_2(\FF_2)$, and since the
elements $(a,0)$ and $(0,b)$ can be identified with $a,b \in \FF_4$
respectively, their image yields an invertible matrix in $\Mc_2(\FF_2)$
whenever $a,b \in \FF_4^*$. These 6 elements thus correspond to the 6
invertible matrices of $\Mc_2(\FF_2)$. We will show below that a similar
correspondence holds for $\Mc_2(\FF_2[i])$ via
$\psi$. The correct phrasing which takes into account both $\FF_4$ and
$\FF_4[i]$ is that there is a correspondence between pairs $(a,a')$ and
$(b',b)$ where $a,b$ are units, while $a',b'$ are not. In the case of
$\FF_4$, $a'=b'=0$, while $a',b'$ are a multiple of $(1+i)$ for $\FF_4[i]$,
as shown below.

\begin{lemma}\label{lem:F4inv}
The ring $\FF_4[i]$ contains exactly 4 non-invertible elements, given by
\[
a(1+i),~a\in\FF_4.
\]
\end{lemma}
\begin{IEEEproof}
Let $x \in \FF_4[i]$, $x=a+ib$, $a,b \in \FF_4$.
\begin{itemize}
\item
if $a=b=0$, $x$ is clearly non invertible.
\item
if $a=0$, $b\neq 0$, then $x=ib$ and $x^3=-i=i$, thus $x$ is invertible.
\item
if $a\neq 0$, $b=0$, then $x=a$ and $a^3=1$, thus $x$ is invertible.
\item
if $a\neq 0$, $b\neq 0$, then $x=a(1+iba^{-1})$ and $x^3=(1+ic)^3$ with
$c\neq 0$.
Now $(1+ic)^2=1-c^2=1+c^2$ and $(1+c^2)^2=1+c$, so that if $c\neq 1$,
$x$ is invertible, and if $c=1$ (that is $a=b$), then $x$ is not invertible.
\end{itemize}
To summarize, $\FF_4[i]$ has 16 elements, 4 of them non invertible (given
by $a(1+i)$, $a\in \FF_4$) and the 12 others being invertible.
\end{IEEEproof}
\begin{prop}
If $(a+b\omega,c+d\omega)\in  \FF_4[i]\backslash\FF_4[i]^* \times\FF_4[i]^*$
(that is, $a+b\omega$ is not invertible and $c+d\omega$ is, or vice versa),
then $\psi((a+b\omega,c+d\omega))$ is invertible.
\end{prop}
\begin{IEEEproof}
Recall by (\ref{eq:iso}) that
\[
(a+b\omega,c+d\omega)
\mapsto
\left(
\begin{array}{cc}
a+d & b+c \\
b+c+d & a+b+d \\
\end{array}
\right).
\]
Since we have assumed that $a+b\omega$ is not invertible, we know by
Lemma \ref{lem:F4inv} that $a+b\omega=a'(1+i)+b'(1+i)\omega$ ($a',b'$
possibly 0). Thus
\[
\psi(a'(1+i)+b'(1+i)\omega,c+d\omega)
=
\left(
\begin{array}{cc}
a'(1+i)+d & b'(1+i)+c \\
b'(1+i)+c+d & a'(1+i)+b'(1+i)+d \\
\end{array}
\right)
\]
whose determinant is $d^2+c(c+d)$.
Since
\[
N_{\FF_4/\FF_2}(c+d\omega)=(c+d\omega)(c+d\omega^2)=c^2+cd+d^2,
\]
the determinant has to be invertible since we have assumed that $c+d\omega$ is
invertible. The vice versa case follows similarly.
\end{IEEEproof}
\begin{cor}
There is a one to one correspondence between ordered pairs
$(a+b\omega,c+d\omega)\in\FF_4[i]\times\FF_4[i]$ formed by one invertible
and one non-invertible element, and invertible matrices in $\Mc_2(\FF_2[i])$.
\end{cor}
\begin{IEEEproof}
There are twice $4\cdot 12$ ordered pairs formed by one invertible and one
non-invertible element, that is $96$ pairs.

On the other hand, let us count invertible matrices in $\Mc_2(\FF_2[i])$.
For the first column, there are a priori 16 choices, from which we have to
remove the following pairs, yielding necessarily non-invertible matrices:
\[
(0,0),(0,1+i),(1+i,0),(1+i,1+i).
\]
That let us 12 choices, 4 choices twice for pairing an invertible ($1$ or $i$)
with a non-invertible ($0$ or $1+i$), and 4 choices for pairing two invertible
elements. We now choose the second column, in such a way that we get an
invertible determinant. For the 8 choices of first columns where there is
an invertible and a non-invertible, the non-invertible can multiply any element
in $\FF_2[i]$, yielding 4 choices, while the invertible is left to be multiplied
by 2 choices. This is thus twice $4\cdot 4\cdot 2 = 4 \cdot 8$ choices.
For the 4 choices with two invertible elements, it is not difficult to see that
in each case, we have 4 choices for the first element of the second column,
and only 2 choices for the second element, for a total of $4\cdot 8$.
Thus the total of invertible matrices is $3\cdot 4\cdot 8=96$.
\end{IEEEproof}

This can be made even more precise.
As in the above proof, we use (\ref{eq:iso}), which holds for $\psi$ and
$a,b,c,d\in\FF_2[i]$ as for $\phi$ and $a,b,c,d\in\FF_2$, to see that
\[
(a+b\omega,c+d\omega)
\mapsto
\left(
\begin{array}{cc}
a+d & b+c \\
b+c+d & a+b+d \\
\end{array}
\right)
\]
and we compute
\begin{eqnarray*}
\det
\left(
\begin{array}{cc}
a+d & b+c \\
b+c+d & a+b+d \\
\end{array}
\right)
&=& (a^2+ab+b^2)+(d^2+cd+c^2)\\
&=& N_{\FF_4[i]/\FF_2[i]}(a+b\omega)+N_{\FF_4[i]/\FF_2[i]}(c+d\omega).
\end{eqnarray*}
Now $N_{\FF_4[i]/\FF_2[i]}(a+b\omega)\in\{0,1,i\}$ and
$N_{\FF_4[i]/\FF_2[i]}(a+b\omega)=0$ when $a+b\omega$ is not invertible, that is,
is a multiple of $1+i$. Thus, we have three different scenarios for
$\det(\phi(a+b\omega,c+d\omega))$:
\begin{itemize}
\item
$N_{\FF_4[i]/\FF_2[i]}(a+b\omega)=N_{\FF_4[i]/\FF_2[i]}(c+d\omega)$: this can happen
either when both $a+b\omega$ and $c+d\omega$ are not invertible, with then
both a norm of zero (no element has norm $1+i$), or both are invertible,
with a norm of either $1$ or $i$.
\item
$N_{\FF_4[i]/\FF_2[i]}(a+b\omega)\neq N_{\FF_4[i]/\FF_2[i]}(c+d\omega)$ with
$N_{\FF_4[i]/\FF_2[i]}(a+b\omega)\neq 0,N_{\FF_4[i]/\FF_2[i]}(c+d\omega)\neq 0$: this means
that $N_{\FF_4[i]/\FF_2[i]}(a+b\omega)=i$, $N_{\FF_4[i]/\FF_2[i]}(c+d\omega)=1$, or
vice-versa.
\item
Either $N_{\FF_4[i]/\FF_2[i]}(a+b\omega)$ or $N_{\FF_4[i]/\FF_2[i]}(c+d\omega)= 0$, thus
the non-zero norm is 1 or $i$.
\end{itemize}

%**************************************************************************%

\subsection{Weights and codes over $\Mc_2(\FF_2[i])$}

First we notice that Lemma \ref{lem:hambound} can easily be restated here:
\begin{lemma}
We have that
\[
\Delta_{min}\geq \min(16\delta,(\dmin^H)^2\delta),
\]
where $\delta=\min|\det(X)|^2$.
\end{lemma}
This gives a Hamming distance bound. We now derive a new bound based on
a bidimensional Lee-like distance.
The determinant of a codeword $X\in\Gc$ can be bounded depending on
its projection
\[
\pi(X)=
\left(
\begin{array}{cc}
a+d & b+c \\
b+c+d & a+b+d \\
\end{array}
\right)
\]
as follows.
\begin{lemma}\label{lem:det2}
We have that
\begin{enumerate}
\item
$|\det(X)|^2\geq 4\delta$ if
$N_{\FF_4[i]/\FF_2[i]}(a+b\omega)=N_{\FF_4[i]/\FF_2[i]}(c+d\omega)$, $X\neq\zom$.
\item
$|\det(X)|^2 \geq 2\delta$ if
$N_{\FF_4[i]/\FF_2[i]}(a+b\omega)\neq N_{\FF_4[i]/\FF_2[i]}(c+d\omega)$ with
$N_{\FF_4[i]/\FF_2[i]}(a+b\omega)\neq 0,N_{\FF_4[i]/\FF_2[i]}(c+d\omega)\neq 0$.
\item
$|\det(X)|^2 \geq \delta$  if $N_{\FF_4[i]/\FF_2[i]}(a+b\omega)$ or
$N_{\FF_4[i]/\FF_2[i]}(c+d\omega)= 0$.
\end{enumerate}
\end{lemma}
\begin{IEEEproof}
\begin{enumerate}
\item
If $N_{\FF_4[i]/\FF_2[i]}(a+b\omega)=N_{\FF_4[i]/\FF_2[i]}(c+d\omega)=0$, then
$\det(\pi(X))=0$, thus $\det(X)$ is a multiple of 2 (assuming $X\neq \zom$).
\item
If $N_{\FF_4[i]/\FF_2[i]}(a+b\omega)\neq N_{\FF_4[i]/\FF_2[i]}(c+d\omega)$ with
$N_{\FF_4[i]/\FF_2[i]}(a+b\omega)\neq 0,N_{\FF_4[i]/\FF_2[i]}(c+d\omega)\neq 0$, then
$\det(\pi(X))=1+i$ and $\det(X)$ is a multiple of $1+i$.
\item
If $N_{\FF_4[i]/\FF_2[i]}(a+b\omega)$ or $N_{\FF_4[i]/\FF_2[i]}(c+d\omega)= 0$, then the
non-zero norm is $1$ or $i$, so that $\det(\pi(X))=1$ or $i$, and
consequently $\det(X)$ is a multiple of $1$ or $i$.
\end{enumerate}
\end{IEEEproof}

The above suggests to define a weight $w_L$ on $(a+bw,c+dw)\in\FF_4[i]^2$ by
looking at their norm as follows:
\[
w_L(N_{\FF_4[i]/\FF_2[i]}(a+b\omega),N_{\FF_4[i]/\FF_2[i]}(c+d\omega))=
|N_{\FF_4[i]/\FF_2[i]}(a+b\omega)+N_{\FF_4[i]/\FF_2[i]}(c+d\omega)|^2\in\ZZ[i],
\]
that is, embed each norm in $\FF_2[i]$ in $\ZZ[i]$ and compute the complex
module of the sum. This can be seen as some bidimensional Lee weight.
It is easy to check that
\begin{eqnarray*}
w_L(1,1) &= w_L(i,i) &= 4\\
w_L(1,i) &= w_L(i,1) &= 2  \\
w_L(0,i) &= w_L(0,1) &= 1
\end{eqnarray*}
as desired.

It is not obvious how to translate this weight defined by norms on
$\FF_2[i]^2$ to $\FF_4[i]^2$. To handle it, we first use an inner code
that will remove the pairs of lowest weights, in order to have only two
weights to distinguish.
We propose to use the inner code given by the parity check matrix
$H=(1+i,1+i)$, that is
\[
(1+i,1+i)
\left(
\begin{array}{c}
a+b\omega\\
c+d\omega
\end{array}
\right)=
(1+i)(a+b\omega+c+d\omega).
\]
Now this parity equation implies that $a+b\omega+c+d\omega$ is
of the form $(1+i)\alpha,\alpha=\alpha_1+w\alpha_2\in\FF_4$, so that pairs satisfying the
parity equation are
\[
(a+b\omega,a+b\omega+\alpha(1+i))\in\FF_4[i]^2,
\]
that is bits encode $a$, 2 bits encode $b$, then 2 more bits
decide which of the 4 multiples $0,1+i,(1+i)w,(1+i)w^2$ is used.
The corresponding matrix is
\[
\left(
\begin{array}{cc}
a+b+\alpha_2(1+i) & b+a+\alpha_1(1+i)\\
a+(1+i)(\alpha_1+\alpha_2)&a+\alpha_2(1+i)
\end{array}
\right).
\]
The rate is consequently $6/8=3/4$ and we have that
\[
w_L(a+b\omega,a+b\omega+\alpha(1+i))
=|a\alpha_2(1+i)+b\alpha_1(1+i)|^2=|1+i|^2||a\alpha_2+b\alpha_1|^2\geq 2,
\]
which is what we wanted, since
\begin{eqnarray*}
&&N_{\FF_4[i]/\FF_2[i]}(a+b\omega+\alpha(1+i))\\
&=& N_{\FF_4[i]/\FF_2[i]}(a+b\omega+\alpha_1(1+i)+\alpha_2w(1+i))\\
&=& (a+\alpha_1(1+i))^2+(b+\alpha_2(1+i))^2+
(a+\alpha_1(1+i))(b+\alpha_2(1+i))\\
&=& a^2+b^2+ab+a\alpha_2(1+i)+b\alpha_1(1+i)
\end{eqnarray*}
and
\[
N_{\FF_4[i]/\FF_2[i]}(a+b\omega)+N_{\FF_4[i]/\FF_2[i]}(a+b\omega+\alpha(1+i))=
a\alpha_2(1+i)+b\alpha_1(1+i).
\]

We are now left with designing a multilevel code, made of an outer code
$C_1$ over $\FF_4[i]$, with minimum Hamming distance $d_1$, and a code
$C_2$ over $\FF_4$ for encoding $\alpha$ with minimum Hamming distance
$d_2$. The total minimum distance is
\[
\min\{2d_1,\sqrt{2}d_2\}.
\]

The goal is to reach a minimum of 4, for which we can take
\begin{itemize}
\item
the parity check code $(L,L-1,2)$  over $\FF_4[i]$,
\item
an $(L,k,d)$ code over $\FF_4$ with the same $L$ and $d\geq 3$.
\end{itemize}
The rate $R$ of the code depends on $L$ and $k$ as follows:
\[
R=\frac{L-1}{2L}+\frac{k}{4L}.
\]
For example, one could take the code $(4,2,3)$ over $\FF_4$ and
$(4,3,2)$ over $\FF_4[i]$, where we choose for $(4,3,2)$ the parity check
code.

For $L\rightarrow\infty$, Gilbert Varshamov bound
\[
A_q(L,d)\geq \frac{q^L}{\sum_{i=0}^{d-1}{L\choose i}(q-1)^i}
\]
predicts that
\[
A_4(L,3)\geq \frac{4^L}{\sum_{i=0}^{d-1}{1+3L+9\frac{L(L-1)}{2}}}
\]
thus
\[
R \geq \frac{1}{2}+\frac{1}{4}\left(1-\frac{2}{L}\log_4{L}+o(1/L)\right).
\]
Therefore the rate satisfies $1\leq R \leq 3/4$.
The normalized redundancy is from (\ref{eq:rnorm})
\begin{eqnarray*}
\rnorm&=&\frac{(1+L)\mbox{ symb in }\FF_4[i]+(L-k)\mbox{ symb in }\FF_4}{2L}\\
      &=&\frac{4(1+L)+2(L-k)}{2L}=\frac{2+3L-k}{L}.
\end{eqnarray*}

\begin{figure}
\begin{center}
\label{fig:m2f2i}
\includegraphics[scale=0.7]{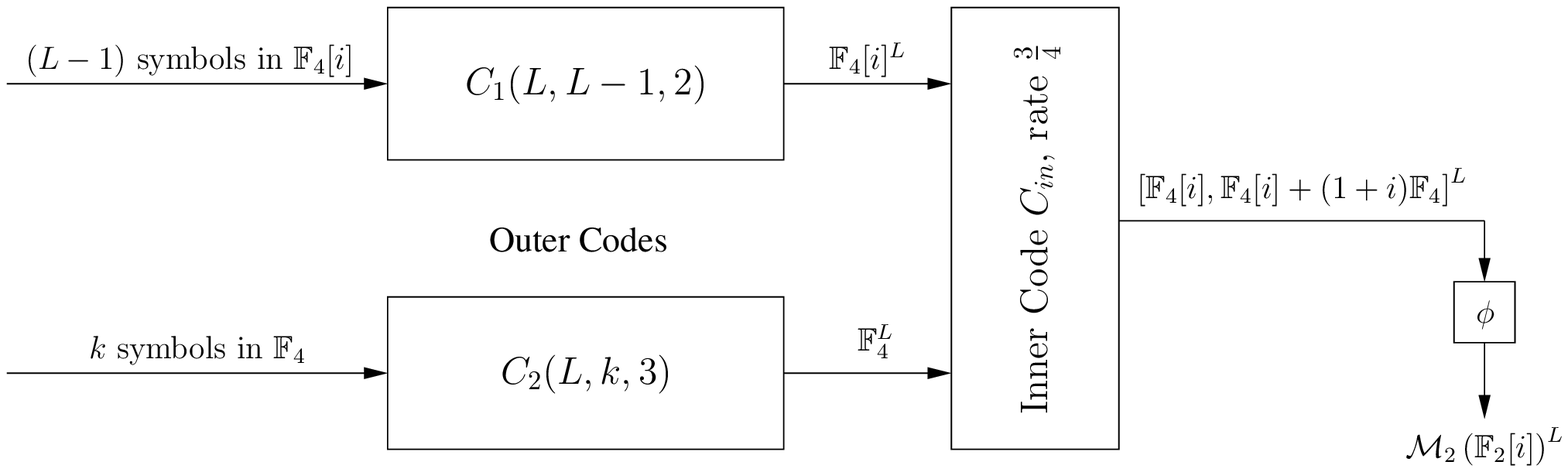}
\caption{Multilevel encoder for $\Mc_2(\FF_2[i])$.}
\end{center}
\end{figure}

%**************************************************************************%
%
% summary
%
%**************************************************************************%

\section{Summary and Perspectives}
\label{sec:summary}
In this paper, we designed coset codes for quasi-static MIMO fading channels where
the inner code comes from a cyclic division algebra. In this case, we showed that the
outer code alphabet is a matrix ring over a finite field or ring of the form
$\Mc_n(\mathcal{R})$ where $n$ is the number of transmit antennas and $\mathcal{R}$
is a finite ring in characteristic $2$.
More precisely, we considered the following cases:
\begin{itemize}
 \item Codes over $\Mc_2(\FF_2)$ with Hamming distance
\item Codes over $\Mc_2(\FF_2)$ with Bachoc distance
 \item Multilevel codes over $\Mc_4(\FF_2)$ with Hamming distance
 \item Multilevel concatenated codes over $\Mc_2(\FF_2[i])$ with a bidimensional Lee-like distance.
\end{itemize}

We established a general framework for designing coset codes via a series of isomorphisms that allows to represent
the outer code alphabet in three different ways: an algebra of matrices over a finite ring,
a cyclic algebra over a finite ring, and the Cartesian product of finite rings.
Under this framework we can address the following scenarios:
\begin{itemize}
 \item For $n=2$, in order to increase the coding gain of the space-time code, we need to consider deeper
levels of partitioning giving rise to larger alphabets $\Mc_2(\mathcal{R})$ with $\mathcal{R}\supset\FF_2[i]$.
 \item For $n=3$, none of the constructions proposed in this paper properly works over $\Mc_3(\FF_4)$.
 \item More generally, one may study deeper levels of partitioning in higher dimensions.
\end{itemize}
For all the above cases, the question of finding a suitable distance and correspondingly designing codes remains open.

%**************************************************************************%
%
% BIBLIO
%
%**************************************************************************%

\end{document}